\documentclass{article}
\usepackage{FPSpro}
\newcommand{\lbl}[1]{\label{eq:#1}}
\newcommand{ \rf}[1]{(\ref{eq:#1})}

\newcommand{\be}{\begin{equation}}
\newcommand{\ee}{\end{equation}}
\newcommand{\bea}{\begin{eqnarray}}
\newcommand{\eea}{\end{eqnarray}}

\newcommand{\noi}{\noindent}

\newcommand{\ra}{\rightarrow}
\newcommand{\Ra}{\Rightarrow}

\newcommand{\lesssim}{ {\
\lower-1.2pt\vbox{\hbox{\rlap{$<$}\lower5pt\vbox{\hbox{$\sim$}}}}\ }
}
\newcommand{\gtrsim}{ {\
\lower-1.2pt\vbox{\hbox{\rlap{$>$}\lower5pt\vbox{\hbox{$\sim$}}}}\ }
}

\newcommand{\cL}{{\cal L}}

\newcommand{\cO}{{\cal O}}

\newcommand{\tr}{\mbox{\rm tr}}

\newcommand{\MeV}{\mbox{\rm MeV}}
\newcommand{\GeV}{\mbox{\rm GeV}}

\newcommand{\annd}{\mbox{\rm and}}

\newcommand{\GF}{G_{\mbox{\rm {\tiny F}}}}

\newcommand{\eff}{\mbox{\rm {\footnotesize}eff}}

\newcommand{\stern}{\langle\bar{\psi}\psi\rangle}

\usepackage{graphicx}
\begin{document}
\title{
 LARGE--$N_C$ METHODS IN KAON PHYSICS
  }
\author{
Eduardo de Rafael      \\
  {\em CPT, CNRS--Luminy, Marseille} \\
  }
\maketitle
\baselineskip=11.6pt
\begin{abstract}
This talk  reviews  recent progress in formulating the dynamics of
Kaon Physics, within the
framework of the $1/N_c$--expansion in QCD.
\end{abstract}
\baselineskip=14pt
\section{Introduction}

In the Standard Model, the electroweak interactions of hadrons at
very low energies can be described by an effective Lagrangian which
only has as active degrees of freedom the flavour $SU(3)$ octet of
the low--lying pseudoscalar particles. The underlying theory,
however, is the gauge theory $SU(3)_{C}\times SU(2)_{L}\times
U(1)_{Y_w}$ which has as dynamical degrees of freedom quarks and
gauge fields. Going from these degrees
of freedom at high energies to an effective description in terms of
mesons at low energies is, in principle, a problem which
should be understood in terms of the evolution of the renormalization
group from short--distances to long--distances. Unfortunately, it is
difficult to carry out explicitly this evolution because at energies,
typically of a few
$\GeV$, non--perturbative dynamics like spontaneous chiral
symmetry breaking and color confinement sets in.

The suggestion to keep the number of colours  $N_c$ in QCD as a free
parameter was first made by G.~'t Hooft\cite{THFT74} as a possible
way to approach the study of non--perturbative phenomena. Many
interesting properties have been proved, which suggest that, indeed, the
theory in this limit has the bulk of the non--perturbative properties of
the full QCD.
The spectrum of the
theory in the large--$N_c$ limit consists of an infinite number of
narrow stable meson states which are flavour nonets\cite{W79}.
This spectrum looks {\it a priori} rather
different to the real world. The vector and axial--vector
spectral functions measured in $e^+ e^- \ra$ hadrons and in the
hadronic
$\tau$--decay show indeed a richer structure than just a sum of
narrow states. There are, however, many instances
where one is only interested in observables which are given by weighted
integrals of some hadronic spectral functions.  Typical examples of that
are the coupling constants of the effective chiral
Lagrangian of QCD at low energies, as well as the coupling constants
of the effective chiral Lagrangian of the electroweak interactions of
pseudoscalar particles in the Standard Model. Some of these
couplings are the ones needed to understand
$K$--Physics quantitatively. In these examples the
{\it hadronic world} predicted by large--$N_c$ QCD  provides already
a good approximation to the real hadronic spectrum. It is in this sense
that I shall show that large--$N_c$ QCD is a very useful
phenomenological approach for understanding non--perturbative QCD
physics at low energies; $K$--Physics in particular.

\section{The Chiral Lagrangian at Low Energies}

Typical terms
of the chiral Lagrangian at low energies are
$$
 \cL_{\eff}   =   \underbrace{\frac{1}4F_{0}^2\
\tr\left(D_{\mu}U
D^{\mu}U^{\dagger}\right)}_{
\pi\pi\ra\pi\pi\,,\quad K\ra\pi e\nu
}+\underbrace{L_{10}
\tr\left(
U^{\dagger}F_{R\mu\nu}
UF_{L}^{\mu\nu}\right)}_{ \pi\ra
e\nu\gamma }+\cdots
$$
\be\lbl{chiral}
\underbrace{e^{2}C
\tr\left(
Q_{R}UQ_{L}U^{\dagger}\right)}_{
-e^2 C\frac{2}{F_{0}^{2}}\left(\pi^{+}\pi^{-}+K^{+}K^{-}
\right)+\cdots}+\cdots -
\underbrace{\frac{\GF}{\sqrt{2}}V_{ud}V^{*}_{us}
\ g_{8}F_{0}^{2}\left(D_{\mu}U
D^{\mu}U^{\dagger}\right)_{23}}_{
{K\ra\pi\pi\,, \quad K\ra\pi\pi\pi\,,\quad+\cdots}}+\cdots \,,
\ee
where $U$ is a $3\times 3$ unitary matrix in flavour space which
collects the fields of the low--lying pseudoscalar particles,
(the Goldstone fields of spontaneous chiral symmetry breaking.) The first
line shows typical terms of the strong interactions in the presence of
external currents\cite{GL85}; the second line shows
typical terms which appear when  photons, $W's$ and
$Z's$  have been integrated out in the presence of the strong
interactions. We show under the braces the typical physical processes
to which each term contributes. Each term is modulated by a constant:
$F_{0}^{2}$, $L_{10}$,... $C$...$g_{8}$... which encode the
underlying dynamics responsible for the appearance of the
corresponding effective term.

There are two
crucial observations concerning the relation of these low energy
constants to the underlying theory, that I
want to emphasize.

\begin{itemize}

\item  The low--energy constants of the
Strong Lagrangian, like
$F_{0}^{2}$ and
$L_{10}$,  are the
coefficients of the \underline{\it Taylor expansion}
of appropriate QCD Green's Functions. For example, with
$\Pi_{LR}(Q^2)$  the correlation function of a
left--current with a right--current in the chiral limit, (where the
light quark masses are neglected,)
the Taylor expansion
\be
-Q^2\Pi_{LR}(Q^2)
\vert_{Q^2\ra
0}=F_{0}^{2}-4L_{10}\ Q^2+
\cO\ (Q^4) \,,
\ee
defines the constants $F_{0}^{2}$ and $L_{10}$.

\item By contrast, the low--energy constants
of the ElectroWeak Lagrangian, like e.g.
$C$ and $g_{8}$, are
\underline{\it integrals} of appropriate QCD Green's Functions. For
example, including the effect of weak neutral currents\cite{KPdeR98},
\be\lbl{pimd}
C=\frac{3}{32\pi^2}\int_{0}^{\infty}dQ^2 \left(1-\frac{Q^2}{Q^2 +
M_{Z}^2}
\right)
\left(-Q^2\Pi_{LR}(Q^2)\right)\,.
\ee
Their evaluation requires  a precise {\it matching} of the {\it
short--distance} and the {\it long--distance} contributions of the
underlying Green's functions.

\end{itemize}

\noi
These observations are completely generic, independently of the
$1/N_c$--expansion. The large--$N_c$ approximation helps, however,
because it restricts the {\it analytic structure} of the Green's
functions in general, and
$\Pi_{LR}(Q^2)$ in particular, to be a sum of poles only; e.g., in
large--$N_c$ QCD,
\be\lbl{largeN}
\Pi_{LR}(Q^2)=\sum_{V}\frac{f_{V}^{2}M_{V}^2}{Q^2+M_{V}^2}-
\sum_{A}\frac{f_{A}^{2}M_{A}^2}{Q^2+M_{A}^2}-\frac{F_{0}^{2}}{Q^2}\,,
\ee
where the sums are, in principle, extended to an infinite number of
states.

There are two types of important restrictions on Green's
functions like $\Pi_{LR}(Q^2)$. One type follows from the fact
that, as already stated above, the Taylor expansion at low euclidean
momenta must match the low energy constants of the strong chiral
Lagrangian.
The other type of constraints follows
from the {\it short--distance  properties} of the
underlying Green's functions which can be evaluated using the {\it
operator product expansion} (OPE) technology. In the large--$N_c$ limit,
this results in a series of algebraic sum rules\cite{KdeR98} which
restrict the coupling constants and masses of the hadronic poles.

\subsection{The Minimal Hadronic Ansatz  Approximation  to
Large--$N_c$ QCD}

In most cases of interest, the Green's functions which govern the
low--energy constants of the chiral Lagrangian are order parameters of
spontaneous chiral symmetry breaking; i.e. they vanish, in the chiral
limit, order by order in the perturbative vacuum of QCD. That implies
that they have a power fall--out in $1/Q^2$ at large--$Q^2$, ( e.g.,
the function
$\Pi_{LR}(Q^2)$ falls as
$1/Q^6$.) That also implies that within a finite radius in the
complex $Q^2$--plane, these Green's functions in the large--$N_c$ limit,
only have a {\it finite number of poles}. The {\it\underline {minimal
hadronic ansatz}} (MHA) approximation is fixed  by the minimum number of
poles required to satisfy the OPE constraints.
In the case of the $\Pi_{LR}(Q^2)$ function, the MHA approximation to the
large--$N_c$ expression in Eq.~\rf{largeN} results in the simple function
\be\lbl{MHA}
-Q^2\Pi_{LR}(Q^2)=F_{0}^{2}\frac{M_{V}^2
M_{A}^2}{(Q^2+M_{V}^2)(Q^2+M_{A}^2)}\,.
\ee
Inserting this in Eq.~\rf{pimd} gives a prediction to the
low--energy constant $C$ which governs the electromagnetic
$\pi^{+}-\pi^{0}\equiv\Delta m_{\pi}$ mass difference, with the
result
\be\lbl{mpimha}
\Delta m_{\pi}=(4.9\pm 0.4)\,\MeV\,,\qquad {\mbox{\rm{
MHA to Large--$N_c$ QCD}}}\,,
\ee
to be compared with the experimental value
\be
\Delta m_{\pi}=(4.5936\pm 0.0005)\,\MeV\,,\qquad {\mbox{\rm
Particle Data Book~\cite{PDG00}}}\,.
\ee

The shape of the function in Eq.~\rf{MHA}, normalized to its value at
$Q^2=0$, is shown in Fig.~1 below, (the continuous red curve.)
Also shown in the same plot is the experimental curve, (the green
dotted curve) obtained from the ALEPH collaboration
data\cite{ALEPH}; as well as
the different shapes predicted by
other analytic approaches.  Let us comment on
them individually.

\vskip 1pc
\begin{figure}
\includegraphics[width=4.5in]{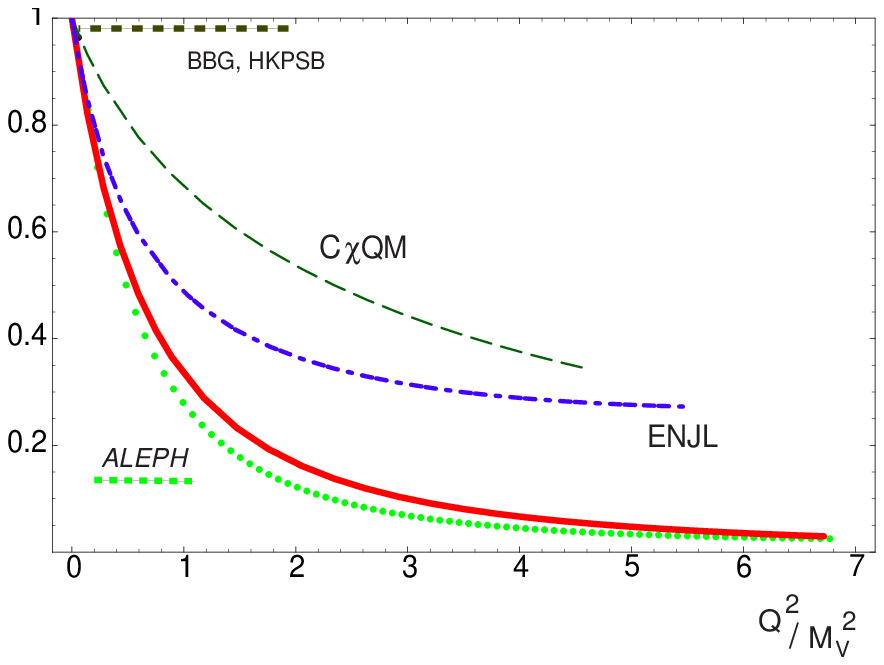}
\caption{MHA (red curve,) versus  ALEPH (green dots) and other
predictions}

\end{figure}

\begin{itemize}
\item The suggestion to use large--$N_c$ QCD combined with lowest order
$\chi$PT loops, was first proposed by Bardeen, Buras and G\'erard in a
series of seminal papers~\footnote{See refs.\cite{Bu88,Ba89,Ge90}
and references therein.}. The same approach has been applied by the
Dortmund group\cite{Dortmmund}, in particular to the evaluation of
$\epsilon'/\epsilon$. In this approach the {\it hadronic ansatz} to
the Green's functions consists of Goldstone poles only. The predicted
hadronic shape, as shown by the BBG, HKPSB line (black dotted), is a
constant.

\item
The Trieste group  evaluate the relevant Green's functions using
the constituent chiral quark model (C$\chi$QM)\cite{MG84,EdeRT90,PdeR91}.
They have obtained a long list of  predictions\cite{Trieste}, in
particular
$\epsilon'/\epsilon$. The model gives an educated first guess of the
low--$Q^2$ behaviour of the Green's functions, as one can judge from
the C$\chi$QM--curve (green dashed), but it
fails to reproduce the short--distance QCD--behaviour.

\item
The extended Nambu--Jona-Lasinio (ENJL) model~\footnote{For a review, see
e.g. ref.\cite{Bij96} where earlier references can be found.}
has a better
low--energy behaviour, as the ENJL--curve
(blue dot--dashed) shows, but it  fails to
reproduce the short--distance behaviour of the OPE in QCD. Arguments to do
the matching to short--distance QCD have been forcefully  elaborated in
refs.\cite{LuGr}, which also have made a lot of predictions; a
large value for
$\epsilon'/\epsilon$ in particular.

\end{itemize}

\noi
In view of the difficulties that these analytic approaches
have in reproducing the shape of  the simplest Green's
function one can think of, it is difficult to attribute more than a
qualitative significance to their ``predictions'';
$\epsilon'/\epsilon$ in particular, which requires the interplay of
several other Green's functions much more complex than
$\Pi_{LR}(Q^2)$.

\section{Applications}

The large--$N_c$ approach that we propose in order to compute a specific
coupling of the chiral electroweak Lagrangian consists of the following
steps:

\begin{enumerate}
\item
{\it Identify the underlying QCD Green's functions.}

\item{\it Work out the short--distance behaviour and the
long--distance behaviour of the relevant Green's functions.}

\item
{\it Make a large--$N_c$ ansatz for the underlying Green's functions.}

\end{enumerate}

\noi
We have tested this approach with the calculation of a few
low--energy observables:

\begin{itemize}
\item  The electroweak $\Delta m_{\pi}$ mass
difference\cite{KPdeR98} which we have already discussed.

\item  The hadronic vacuum polarization contribution to the
anomalous magnetic moment of the muon
$a_{\mu}$\cite{PdeR}, with the result
\be
a_{\mu}\vert_{HVP}=(5.7\pm 1.7)\times 10^{-8}\,,
\ee
to be compared with an average of recent
phenomenological determinations~\footnote{See e.g. Prades's talk at
KAON2001~\cite{Pr01} and references therein.}
\be
a_{\mu}\vert_{HVP}=(6.949\pm 0.064)\times 10^{-8}.
\ee

\item
The $\pi^{0}\ra e^{+}e^{-}$ and $\eta\ra \mu^{+}\mu^{-}$ decay
rates\cite{KPPdeR99}, with the results shown in Table~1 below, where
\be\lbl{rat}
R(P\ra l^{+}l^{-})=\frac{\Gamma(P\ra l^{+}l^{-})}{\Gamma(P\ra
\gamma\gamma)}\,.
\ee

\end{itemize}

\begin{table}[h]
\caption{\em Summary of branching ratios results}
\begin{tabular}{@{}l@{}l@{}l@{}}
\hline\hline
{\bf Branching Ratio}~\rf{rat}~~~ & {\bf Large--$N_c$ Approach}
~~~~~~~~~~ &
{\bf Experiment}\cite{PDG00}\\
\hline
$R(\pi^0\to e^+e^-)\times 10^{8}\qquad$ & $6.2\pm 0.3\qquad$ &
$6.28\pm 0.55$
  \\ \hline
$R(\eta\to \mu^+\mu^-)\times 10^{5}$ & $1.4\pm 0.2$ & $1.47\pm 0.20
$ \\ \hline
$R(\eta\to e^+e^-)\times 10^{8}$ & $1.15\pm 0.05$ &
 $< 1.8\times 10^{4}$
\\
\hline\hline
\end{tabular}
\end{table}

\noi
These successful predictions have encouraged us to start a
project of a systematic analysis of non--leptonic $K$--decays within
this large--$N_c$ approach.
So far we have completed two calculations of $K$--matrix elements
within this large--$N_c$ approach, which we next discuss.

\subsection{The $B_{K}$--Factor of $K^{0}-\bar{K}^{0}$ Mixing}

The factor in question is conventionally defined by the matrix
element of the four--quark operator $Q_{\Delta S=2}(x)=
(\bar{s}_{L}\gamma^{\mu}d_{L})
(\bar{s}_{L}\gamma_{\mu}d_{L})(x)$:
\be
\langle \bar{K}^{0}\vert Q_{\Delta S=2}
(0)\vert
K^{0}\rangle =\frac{4}{3}f_{K}^2 M_{K}^2 B_{K}(\mu)\,.
\ee
To lowest order in the chiral expansion the operator  $Q_{\Delta
S=2}(x)$ bosonizes into a term of $O(p^2)$
\be
Q_{\Delta S=2}(x)\Ra -\frac{F_{0}^4}{4}g_{\Delta S=2}(\mu)
\left[(D^{\mu}U^{\dagger})U\right]_{23}
\left[(D_{\mu}U^{\dagger})U\right]_{23}\,,
\ee
with $g_{\Delta S=2}(\mu)$ a low energy constant, to be determined,
which is a function of the renormalization scale $\mu$ of the
Wilson coefficient $C(\mu)$ which modulates the operator $Q_{\Delta
S=2}(x)$ in the four--quark effective Lagrangian.
The coupling constant $g_{\Delta
S=2}(\mu)$, which has to be evaluated in the same renormalization
scheme as the Wilson coefficient $C(\mu)$ has been calculated, is
given by an integral\cite{PdeR00}
conceptually similar to the one which determines the electroweak
constant
$C$ in Eq.~\rf{pimd}.  The invariant
$\hat{B}_{K}$  defined as
\be
\hat{B}_{K}=\frac{3}{4}C(\mu)\times g_{\Delta S=2}(\mu)\,,
\ee
can then be evaluated, with no free parameters, with the
result\cite{PdeR00}
\be\lbl{bk}
\hat{B}_{K}=0.38\pm 0.11\,.
\ee
When comparing this result to other determinations, specially in
lattice QCD, it should be realized that the unfactorized
contribution to this result is the one in the chiral limit.
It is possible, in principle, to calculate chiral corrections within
the same large--$N_c$ approach, but this has not yet been done.

The result in Eq.~\rf{bk} is compatible with
the old current algebra prediction\cite{DGH82} which,
to lowest order in
chiral perturbation theory, relates the
$B_{K}$ factor to the $K^{+}\ra \pi^{+}\pi^{0}$ decay rate.
In fact, our
calculation of the $B_{K}$ factor can be viewed as a
successful prediction of
the $K^{+}\ra \pi^{+}\pi^{0}$ decay rate!

As discussed in\cite{PdeR96}
the bosonization of the four--quark operator
$Q_{\Delta S=2}$ and the bosonization of the operator
$Q_{2}-Q_{1}$ which generates $\Delta I=1/2$ transitions
are related to each
other in the combined chiral limit and  next--to--leading
order in the $1/N_c$--
expansion. Lowering the value of $\hat{B}_{K}$ from the leading
large--$N_c$ prediction $\hat{B}_{K}=3/4$ to the result in
Eq.~\rf{bk} is
correlated with an increase of the coupling constant $g_{8}$
in the lowest order
effective chiral Lagrangian, (see Eq.~\rf{chiral},) which generates
$\Delta I=1/2$ transitions, and
provides a first step towards a quantitative understanding of
the dynamical
origin of the  $\Delta I=1/2$ rule.

\subsection{ ElectroWeak Four--Quark Operators}

These are the four--quark operators generated by the so called electroweak
Penguin like diagrams~\footnote{See e.g., Buras lectures~\cite{Bu99}}
\be
\cL\Ra \cdots
C_{7}(\mu)Q_{7}+
C_{8}(\mu)
Q_{8}\,,
\ee
with
\be
Q_7 = 6(\bar{s}_{L}\gamma^{\mu}d_{L})
\sum_{q=u,d,s} e_{q} (\bar{q}_{R}\gamma_{\mu}q_{R})
\quad\annd\quad Q_8 =
-12\sum_{q=u,d,s}e_{q}(\bar{s}_{L}q_{R})(\bar{q}_{R}d_{L})\,,
\ee
and $C_{7}(\mu)$, $C_{8}(\mu)$ their corresponding Wilson coefficients.
They generate a term of $O(p^0)$ in the effective chiral
Lagrangian~\cite{BW84}; therefore, the matrix elements of
these operators, although suppressed by an $e^2$ factor, are chirally
enhanced. Furthermore, the Wilson coefficient $C_{8}$ has a large
imaginary part, which makes the matrix elements of the
$Q_{8}$ operator to be particularly important in the evaluation of
$\epsilon'/\epsilon$.

Within the large--$N_c$ framework, the bosonization of these
operators produce matrix elements with the following counting
\be\lbl{Q7Q8z}
\langle Q_{7}\rangle\vert_{O(p^0)}= \underline{O(N_c)}
+O(N_c^0)
\quad\annd\quad
\langle Q_{8}\rangle\vert_{O(p^0)}= \underline{O(N_{c}^2)}
+\!\!\!\!\!\!\!\!\!
\underbrace{O(N_c^0)}_{{\mbox{\rm
Zweig suppressed}}}
\ee
A first estimate of the underlined contributions was made in
ref.\cite{KPdeR99}. The inclusion of final state interaction
effects based on the leading large--$N_c$ determination of $\langle
Q_{8}\rangle$, (and $\langle
Q_{6}\rangle$,) in connection with a phenomenological determination of
$\epsilon'/\epsilon$, has been recently discussed in\cite{PP01}.

The bosonization of the
$Q_{7}$ operator to
$O(p^0)$ in the chiral expansion and to $O(N_c)$ is very similar to
the calculation of the $Z$--contribution to the coupling constant $C$
in Eq.~\rf{pimd}. An evaluation which also takes into account the
renormalization scheme dependence has been recently made
in\cite{KPdeR01}.

The contribution of $O(N_c^0)$ to
$\langle Q_{8}\rangle\vert_{O(p^0)}$ is
Zweig suppressed. It involves the sector of
scalar (pseudoscalar)
Green's functions  where it is hinted from various
phenomenological sources that
the restriction to just the
leading large--$N_c$ contribution may not always be a good
approximation. Fortunately, as first pointed out in\cite{DG00},
independently of large--$N_c$ considerations, the bosonization of the
$Q_8$ operator to $O(p^0)$ in the chiral expansion
can be related to the four--quark condensate
$\langle O_2 \rangle\equiv
\langle 0\vert (\bar{s}_{L}s_{R})(\bar{d}_{R}d_{L})\vert
0\rangle$ by current
algebra Ward identities, the same four--quark condensate which also
appears in the OPE of the $\Pi_{LR}(Q^2)$ function  discussed above.
The crucial observation, here, is that large--$N_c$ QCD gives a rather
good description of the $\Pi_{LR}(Q^2)$--function, as we have seen
earlier; in particular it implies that\cite{KdeR98}
\be
\lim_{Q^2\ra\infty} \left(
-Q^2\Pi_{LR}(Q^2)\right)Q^4=\sum_V f_V^2 M_V^6 - \sum_A f_A^2 M_A^6 \,.
\ee
This results in a
determination of the matrix elements of
$\langle Q_{8}\rangle\vert_{O(p^0)}$ which does not require the
separate knowledge of the Zweig suppressed $O(N_c^{0})$ term in
Eq.~\rf{Q7Q8z}.

The numerical results we get for the matrix elements
\be\lbl{M78}
M_{7,8}\equiv \langle (\pi\pi)_{I=2}\vert Q_{7,8}\vert
K^{0}\rangle_{~2\,
\GeV}
\ee
at the renormalization scale $\mu=2~\GeV $ in the two schemes NDR
and HV and in units of
$\GeV^{3}$ are shown in Table~2 below, (the first line.)
\begin{table}[h]
\caption{\em Matrix elements  results, (see Eq.~\rf{M78})}
\begin{tabular}{@{}l@{}l@{}l@{}l@{}l@{}}
\hline\hline
METHOD & $M_{7}$(NDR)~~~~~~ & $M_{7}$(HV)~~~~~~~ &
$M_{8}$(NDR)~~~~~~~ &
$M_{8}$(HV)~~~~~~\\
\hline
{\bf Large--$N_c$ Approach} & & & & \\
Ref.\cite{KPdeR01} & $0.11\pm 0.03$ & $0.67\pm 0.20$ & $3.5\pm 1.1$
& $3.5\pm 1.1$ \\
\hline
{\bf Lattice QCD} & & & & \\
Ref.\cite{Donetal} & $0.11\pm 0.04$ & $0.18\pm 0.06$ & $0.51\pm 0.10$
& $0.62\pm 0.12$ \\
\hline
{\bf Dispersive Approach} & & & & \\
Ref.\cite{DG00}~~~~~~ & $0.22\pm 0.05$ &  & $1.3\pm 0.3$
& \\
Ref.\cite{Na01} & $0.35\pm 0.10$ &  & $2.7\pm 0.6$ &  \\
Ref.\cite{Go01}~~~~~~~~~& $0.18\pm 0.12$~~~~~&
$ 0.50\pm 0.06 $~~~~~&
$2.13\pm 0.85$~~~~~~~ & $2.44\pm 0.86$ \\
Ref.\cite{CDGM01} & $0.16\pm 0.10$ & $0.49\pm 0.07$ & $2.22\pm 0.67$ &
$2.46\pm 0.70$  \\
\hline\hline
\end{tabular}
\end{table}

\noi
Also shown in the same table
are other
evaluations of matrix elements with which we can compare scheme
dependences
explicitly~\footnote{There is a new "dispersive determination" in the
literature\cite{BGP01} since the KAON2001 conference, but it is
controversial as yet; this is why we do not include it in the Table. }.
Several remarks are in order

\begin{itemize}

\item  Our evaluations of $M_8$ do not include the terms of
$O(\alpha_{s}^2)$ because, as pointed out in
Ref.\cite{KPdeR01}, the available results in the literature\cite{LSC86}
were not calculated in the right basis of four--quark operators needed
here.

\item We find
that our results for $M_{7}$ are in very good
agreement with the lattice results in the NDR scheme, but not in the HV
scheme. This disagreement is, very likely, correlated with the strong
discrepancy we have with the lattice result for $M_{8}$(NDR).

\item
The recent revised dispersive approach results\cite{Go01,CDGM01}, which
now include the  effect of higher
terms in the  OPE, are in agreement, within errors, with
the large--$N_c$ approach results. In fact, the agreement improves further
if the new
$O(\alpha_{s}^2)$ corrections, which have now been calculated in the right
basis\cite{CDGM01}, are also incorporated in the large--$N_c$ approach.

\item
Both the revised dispersive approach results and the
large--$N_c$ approach
results for $M_{8}$ are higher than the lattice results. The
discrepancy may originate in the fact that, for reasonable
values of $\stern$,
most of the contribution to  $M_{8}$ appears to come from an OZI--violating
Green's function which is something inaccessible in the quenched
approximation at which the lattice results, so far, have been obtained.
\end{itemize}

\section{Acknowledgements}
I wish to thank Marc Knecht, Santi Peris, Michel
Perrottet, and Toni Pich for enjoyable collaborations on the various
topics reported here.

\newpage

\noindent
\begin{center}
{\bf Discussion}
\end{center}
\vskip 0.5 cm

{\em Martinelli} - Don't you think that your results for the $M_{8}$
matrix elements could be affected by chiral corrections?

{\em de Rafael} - There are chiral corrections, of course, which so far
have not been evaluated; but the results that you reported in your talk,
showing that the lowest order chiral Ward identities for these matrix
elements are well reproduced in lattice QCD, are an indication that the
chiral corrections should be rather small in this case.

{\em Isidori} -  Does your calculation of the $P \ra \gamma^* \gamma^*$
amplitude, used to predict $P\ra ll$ widths, reproduce well the
available data on the $P\ra\gamma^*\gamma$ form factor?

{\em de Rafael} -  Yes, but I suggest you to discuss this issue with
Andreas Nyffeler,  who has recently performed a detailed study
of this subject with Marc Knecht in hep-ph/0106034.

{\em Roberts} - Could your large--$N_c$ methods be applied to a
calculation of the hadronic light--by--light contribution to $g_{\mu}-2$?

{\em de Rafael} - Yes, and in fact we have been discussing quite a lot
about this in Marseille. We think that it should be possible to do a
rather clean evaluation of the contribution which comes from two $\langle
PVV\rangle$ three--point functions.

{\em Buras} - I am puzzled by the fact that your {\it hadronic}
integrals go all the way to infinity.

{\em de Rafael} - This is due to the fact that we both are using  a
renormalization scheme based on dimensional regularization: you at {\it
short--distances} and we at {\it long--distances}. We both depend on
a scale $\mu$ which has to cancel in the matching. If we were doing a
cut--off renormalization, the {\it long--distance}
integrals would then be cut at a certain scale $\Lambda$ indeed; but then
you would have to redo all the short--distance analyses in the same
cut--off renormalization which, as Donoghue pointed out, would introduce
operators of higher dimension at short--distances.

\end{document}